# Kubo-Martin-Schwinger relation for an interacting mobile impurity

Oleksandr Gamayun,[1, *] Miłosz Panfil,[2] and Felipe Taha Sant'Ana[3]

[1]*London Institute for Mathematical Sciences, Royal Institution, 21 Albemarle St, London W1S 4BS, UK*
[2]*Faculty of Physics, University of Warsaw, ul. Pasteura 5, 02-093 Warsaw, Poland*
[3]*Institute of Physics, Polish Academy of Sciences, Aleja Lotników 32/46, 02-668 Warsaw, Poland*
(Dated: August 22, 2023)

In this work we study the Kubo-Martin-Schwinger (KMS) relation in the Yang-Gaudin model of an interacting mobile impurity. We use the integrability of the model to compute the dynamic injection and ejection Green's functions at finite temperatures. We show that due to separability of the Hilbert space with an impurity, the ejection Green's in a canonical ensemble cannot be reduced to a single expectation value as per microcanonical picture. Instead, it involves a thermal average over contributions from different subspaces of the Hilbert space which, due to the integrability, are resolved using the so-called spin rapidity. It is then natural to consider the injection and ejection Green's functions within each subspace. By means of reformulating the original KMS condition as a Riemann-Hilbert problem, we analytically demonstrate that such Green's functions obey a refined analogous relation, which is finally corroborated by numerical evaluation.

The notion of thermal equilibrium is one of the fundamental concepts of physics. In quantum many-body systems, thermal states are described by the density matrix $\hat\rho = e^{-\beta \hat H}/\mathcal{Z}$ with $\mathrm{tr}\hat\rho = 1$, where $\hat H$ is the Hamiltonian of the system. The same operator $\hat H$ is responsible for the time evolution of the system. This double role of the Hamiltonian is crucially important in establishing the Kubo-Martin-Schwinger [1–3] (KMS) condition between the Green's functions

$$\mathrm{tr}\left(\hat\rho A(t-i\beta)B(0)\right) = \mathrm{tr}\left(\hat\rho B(0)A(t)\right), \qquad (1)$$

where $A, B$ are operators and $A(t)$ follows the Heisenberg evolution $A(t) = e^{i\hat H t} A e^{-i\hat H t}$. The KMS condition involves analytic continuation of the Green's function $\mathrm{tr}(\hat\rho A(t)B(0))$ to imaginary times in a strip $0 < \mathrm{Im}\, t < \beta$. Such a relation is central to the concept of thermal equilibrium. For example, the thermal density matrix $\hat\rho$ can be actually defined as the one for which the relation holds [3–5]. Moreover, the KMS condition is an example of a detailed balance relation which guarantees stability of thermal equilibrium under fluctuations [6–8] and is also a foundation for fluctuation-dissipation relations as established in the original works [1, 2]. Finally, the relation can also be promoted to higher point functions [9].

Whereas the KMS condition simply follows from cyclicity of the trace, it is generally very difficult to show this relation explicitly by evaluating thermal two-point functions in interacting many-body systems and comparing both sides of the equality. The aim of our work is to demonstrate such computations for an impurity Green's functions.

The thermal expectation values appearing in (1) involve averaging over different eigenstates. In the thermodynamically large system and under the equivalence between the grand canonical (GCE) and microcanonical (MCE) ensembles, they can be computed over a single eigenstate,

$$\mathrm{tr}\left(\hat\rho A(t)B(0)\right) = \langle\rho|A(t)B(0)|\rho\rangle, \qquad (2)$$

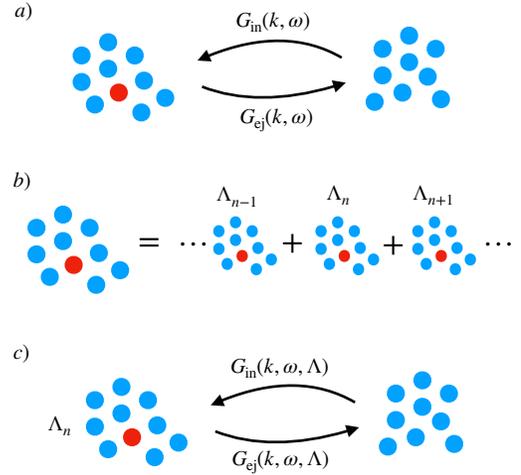

FIG. 1. a) The standard KMS relation infers the detailed balance condition $G_{\mathrm{in}}(k,\omega) = e^{-\beta\omega} G_{\mathrm{ej}}(k,\omega)$, between the Fourier transforms of injection and ejection Green's function, which implies that probabilities for fluctuations creating or destroying impurity (denoted by a red circle) are the same in the thermal equilibrium. b) The thermal state with an impurity is an ensemble of different states enumerated by $\Lambda_n$. c) The refined KMS relation guarantees the stability of each sub-ensemble under the fluctuations creating or destroying the impurity described by $G_{\mathrm{in,ej}}(k,\omega,\Lambda)$.

where the eigenstate $|\rho\rangle$ is chosen such that its energy equals the average energy in the canonical ensemble. Our results show that unexpectedly the thermal average reduces to a single expectation value on one side of the KMS relation but not on the other. This makes the relation itself highly non-trivial. Without the trace, there is no simple derivation of it and it must rely on a more subtle structure of the thermal expectation values which we unravel.

In this work we study the 1d Yang-Gaudin gas of inter-acting fermions. We consider a situation with an extreme population imbalance between spin up and down particles. Such system can be viewed as a mobile impurity (e.g. spin down) immersed in a host medium made of spin up particles. The physics of low dimensional impurity systems has been an active research area [10–26] due to the advances in their experimental realizations [27–30].

The two important correlation functions for the impurity problems are injection and ejection Green's function (we define them precisely below). The two Green's functions are related by the KMS condition, see Fig. 1. However, as we show, the equivalence between GCE and MCE holds only for the injection Green's function and it is broken for the ejection. We attribute its breaking to the Hilbert space separability — the Hilbert space of states with a single impurity separates into different sectors — with each sector contributing a different value to the thermal correlator. Then the latter cannot be reconstructed by a single expectation value. Interestingly, we can define the injection and ejection Green's function between individual sectors of the Hilbert space. As we will show later, this refined Green's function obey a refined KMS relation.

The separability of the Hilbert space in the Yang-Gaudin model is ultimately related to its integrability. However, a similar mechanism called Hilbert space fragmentation [31, 32] appears more generically and leads to breaking of the Eigenstate Thermalization Hypothesis describing the thermalization of closed quantum manybody systems [33–39].

*The model and its correlation functions:* The Yang-Gaudin model [40] describes a 1d gas of spinful fermions with spin polarized interactions, The Hamiltonian density is

$$\mathcal{H} = \sum_{\sigma=\uparrow,\downarrow} -\psi_\sigma^\dagger(x)\partial_x^2\psi_\sigma(x) + g\psi_\uparrow^\dagger(x)\psi_\uparrow(x)\psi_\downarrow^\dagger(x)\psi_\downarrow(x), \tag{3}$$

and we consider repulsive interactions, $g > 0$. The total number of particles of each spin, $N_\sigma = \int dx \psi_\sigma^\dagger(x)\psi_\sigma(x)$ is a constant of motion and the Hamiltonian can be diagonalized in a subspace with fixed numbers of particles of the two kinds. The relevant for the impurity problem are then subspaces with $N$ spin up particles and $0$ or $1$ spin down particle. We denote them by $\mathcal{H}_0$ and $\mathcal{H}_1$ respectively. The dynamics in $\mathcal{H}_0$ is then this of a free spinless Fermi gas, in $\mathcal{H}_1$ it is of the free spinless Fermi gas with a single impurity.

We define two equilibrium dynamic Green's functions of the impurity

$$G_{\text{in}}(x,t) = \text{tr}_0\left(e^{-\beta(H-\mu N_\uparrow)}\psi_\downarrow(x,t)\psi_\downarrow^\dagger(0)\right), \tag{4}$$

$$G_{\text{ej}}(x,t) = \text{tr}_1\left(e^{-\beta(H-\mu N_\uparrow)}\psi_\downarrow^\dagger(x,t)\psi_\downarrow(0)\right), \tag{5}$$

where the trace is over either $\mathcal{H}_0$ or $\mathcal{H}_1$. We note that these are not normalized Green's functions [41]. To restore proper normalization one needs to divide them by the partition function $\mathcal{Z}_i = \text{tr}_i e^{-\beta(H-\mu N_\uparrow)}$. The two (normalized) Green's functions describe response of the system to injection (in) and ejection (ej) of the impurity respectively and can be measured in spectroscopy experiments [42–46]. The KMS relation between the two Green's function was derived in [47, 48] and in our notation takes the following form

$$G_{\text{ej}}(x,t) = \frac{e^{\beta\mu}}{\mathcal{Z}_{\text{ej}}}G_{\text{in}}(-x,-t-i\beta). \tag{6}$$

The extra factor can be seen as a ratio of $\mathcal{Z}_0/\mathcal{Z}_1$ defined for the Green's functions (4), (5).

After this introduction we can now state our results. These confirm the microcanonical picture for the injection function and disprove it for the ejection function. In the former case we find

$$G_{\text{in}}(x,t) = \langle\rho|\psi(x,t)\psi^\dagger(0)|\rho\rangle, \tag{7}$$

with $|\rho\rangle$ a representative state of the thermal equilibrium for free fermionic gas. Instead, for the *ejection* function there is a single remaining degree of freedom, the rapidity $\Lambda$, which is related to the way the impurity and the gas share the total momentum in the system,

$$G_{\text{ej}}(x,t) = \frac{1}{\mathcal{Z}_{\text{ej}}}\int \frac{d\Lambda}{2\pi} e^{-\beta\mathcal{F}(\Lambda)} G_{\text{ej}}(x,t,\Lambda), \tag{8}$$

with the free energy $\mathcal{F}(\Lambda)$ associated with varying $\Lambda$, the corresponding partition function $\mathcal{Z}_{\text{ej}}$, and the $\Lambda$-resolved ejection function $G_{\text{ej}}(x,t,\Lambda)$. Whereas the microcanonical picture does not work for the whole correlator, once the value of $\Lambda$ is fixed, the thermal average reduces to a single expectation value,

$$G_{\text{ej}}(x,t,\Lambda) = \langle\rho_\Lambda|\psi^\dagger(x,t)\psi(0)|\rho_\Lambda\rangle, \tag{9}$$

with $|\rho_\Lambda\rangle$ denoting a representative thermal eigenstate with fixed value of $\Lambda$. This is similar to the generalized ETH [49, 50] appearing in the equilibration processes of integrable models. There, the expectation value can be represented by a single eigenstate once *all* the conserved charges are fixed. Here, it is sufficient to fix a single additional degree of freedom $\Lambda$.

This observation is formalized by realizing the Hilbert space with a single impurity through a direct sum of subspaces. In a finite system, values of $\Lambda$ are quantized and infinite. Denoting the possible values by $\Lambda_m$ with $m \in \mathbb{Z}$ we can formally write

$$\mathcal{H}_1 = \cdots \oplus \mathcal{H}_1(\Lambda_{-1}) \oplus \mathcal{H}_1(\Lambda_0) \oplus \mathcal{H}_1(\Lambda_1) \oplus \ldots. \tag{10}$$

This structure allows us to formally define the projection operator in different subspaces denoted $\mathcal{H}_1(\Lambda)$,

$$P_\Lambda : \mathcal{H}_1 \to \mathcal{H}_1(\Lambda), \quad P_\Lambda^2 = P_\Lambda. \tag{11}$$

The projection operator can be now used to define a $\Lambda$-resolved injection function

$$G_{\text{in}}(x,t,\Lambda) = \langle \rho | \psi(x,t) P_\Lambda \psi^\dagger(0) | \rho \rangle, \quad (12)$$

such that

$$G_{\text{in}}(x,t) = \int \frac{d\Lambda}{2\pi} G_{\text{in}}(x,t,\Lambda). \quad (13)$$

The two $\Lambda$-resolved correlation functions obey a refined KMS condition

$$G_{\text{ej}}(x,t,\Lambda) = e^{\beta(\mu+\mathcal{F}(\Lambda))} G_{\text{in}}(-x,-t-i\beta,\Lambda). \quad (14)$$

Prove of this relation is the main result of work. This new KMS relation implies that $\Lambda$ acquires a thermodynamic meaning, its refines the concept of thermal equilibrium to be specified not only by the temperature and the chemical potential $\mu$ but also by the rapidity $\Lambda$. The $\Lambda$-resolved KMS relation implies then the stability of this generalized equilibrium state. Integrating (14) over $\Lambda$ provides an alternative to [47, 48] derivation of the KMS relation (6).

*Bethe ansatz solution to the McGuire model:* We present now the relevant ingredients of the Bethe ansatz solution to the Yang-Gaudin model with a single spin down and refer to [51] for details. This special case is known as McGuire model [52, 53]. The system's eigenstates $|\{k_j\},\Lambda\rangle$ are specified by a set of (N+1) rapidities $\{k_j\}$ together with an extra rapidity $\Lambda$. For a system of length $L$ with periodic boundary conditions the rapidities $k_j$ are solutions to the Bethe equations

$$k_j = \frac{2\pi}{L}\left(n_j - \frac{\delta(k_j)}{\pi}\right), \quad j=1,\ldots,N+1 \quad (15)$$

where quantum numbers $n_j$ are integers and obey the Pauli principle and the phase shift is

$$\delta(k) = \frac{\pi}{2} - \arctan(\Lambda - \alpha k), \quad \alpha = \frac{2\pi}{g}\frac{N}{L}. \quad (16)$$

The allowed values for the rapidity $\Lambda$ are obtained by requiring that the total momentum is quantized

$$P \equiv \sum_j k_j = \frac{2\pi I}{L}, \quad I \in \mathbb{Z}. \quad (17)$$

Therefore, the state of the system is characterized by a choice of quantum numbers $\{n_j\}$ and $I$ and the rapidities $\Lambda$ and $\{k_j\}$ follow from the Bethe equations. Because the choice of the quantum number $I$ is independent of the other quantum numbers the whole Hilbert space separates into subspaces with fixed value of $I$, or equivalently, with fixed value of $\Lambda$ as anticipated in (10). In a large system there is then a density of states $a(\Lambda)$ associated to a given $\Lambda$,

$$a(\Lambda) = \frac{\partial P}{\partial \Lambda} = \int \frac{dk}{\pi} \frac{\sigma(k)}{1+(\alpha k - \Lambda)^2}, \quad (18)$$

where $\sigma(k) = 1/(1+e^{\beta(k^2/2-\mu)})$ is the Fermi-Dirac distribution. The density of states which is expected to appear in (8) and (13) is absorbed in the definition of the $\Lambda$-resolved Green's functions. Finally, the impurity free energy is [54, 55]

$$\mathcal{F}(\Lambda) = -2\int \frac{dk}{2\pi} k\sigma(k)\delta(k). \quad (19)$$

*Impurity Green's functions:* The program of computations of impurity Green's functions was initiated in [56] where the zero temperature static injection Green's function was computed. This was subsequently generalized to dynamic correlators [57] and to finite temperatures [58]. Similar techniques were later applied to determine the momentum distribution function of the impurity in a zero-temperature polaron state [59]. On the other hand, the static finite temperature ejection Green's function was computed in [55]. This can be then generalized to time-dependent Green's function as we show in the Supplementary Materials [60]. This approach culminates in the following expressions for the two Green's functions:

$$G_{\text{in}}(x,t,\Lambda) = \det_\sigma(1+\hat{V}-\hat{W}_+) - H\det_\sigma(1+\hat{V}), \quad (20)$$

$$G_{\text{ej}}(x,t,\Lambda) = \det_\sigma(1+\hat{V}-\hat{W}_-) - \det_\sigma(1+\hat{V}). \quad (21)$$

Here $\det_\sigma(1+\hat{K})$ denotes a Fredholm determinant of the kernel $K(q,q')$ with a measure given by the Fermi-Dirac distribution $\sigma(k)$ [61]. The kernels appearing above take the following universal form

$$V(q,q') = \frac{1}{\pi}\frac{e_+(q)e_-(q')-e_+(q')e_-(q)}{q-q'}, \quad (22)$$

$$W_\pm(q,q') = \pm\frac{1}{\pi}e_\pm(q)e_\pm(q'), \quad (23)$$

where $e_+(q) = e(q)e_-(q)$. Functions $e_-(q)$ and $e(q)$ have different expressions for the injection and ejection. Namely, for the injection we have

$$e_-^{\text{in}}(q) = e^{itq^2/4-ixq/2} \quad (24)$$

$$e^{\text{in}}(q) = \frac{F(q+i0)-F(\kappa_+)}{\alpha q - \Lambda - i} - \frac{F(q-i0)-F(\kappa_-)}{\alpha q - \Lambda + i}, \quad (25)$$

and additionally $H = 1 - F(\kappa_+)/\alpha + F(\kappa_-)/\alpha$. Here $\kappa_\pm = (\Lambda \pm i)/\alpha$ and

$$F(z) = \int \frac{dk}{2\pi i}\frac{e^{-itk^2/2+ikx}}{k-z}. \quad (26)$$

For a real argument we specify how we deform a contour to pass over or under a real line by $\pm i0$ shifts.

For the ejection the formulas read

$$e_-^{\text{ej}}(q) = e^{itq^2/4-ixq/2}\frac{1}{\sqrt{1+(\Lambda-\alpha q)^2}}, \quad (27)$$

$$e^{\text{ej}}(q) = 2iF(q-i0) - e^{-itq^2/2+ixq}(\alpha q - \Lambda - i). \quad (28)$$

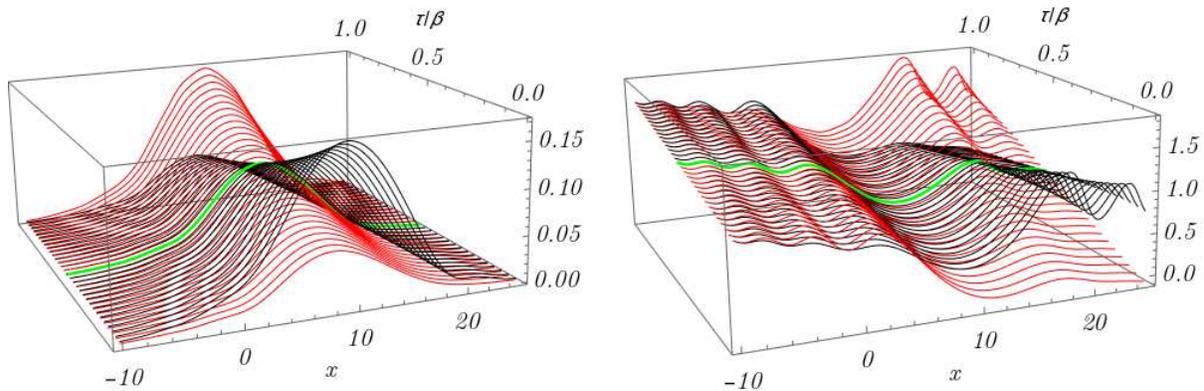

FIG. 2. The KMS relation between the impurity's Green's functions for $\alpha = 1$, $\Lambda = 1$, $\beta = 2$, $\mu = 1/2$. Black lines represent $G_{\text{in}}(x, t_0 - i\tau)e^{-ik_0 x}$ red lines $e^{-\beta\mu - \beta\mathcal{F}_0} G_{\text{ej}}(-x, -t_0 - i\tau)e^{-ik_0 x}$. For fixed vales $t_0 = 10$ and $k_0 = 0.63$ (non-zero $k_0$ is chosen for the demonstration purposes). The left panel shows absolute values and the right panel arguments. The green line shows a cut $\tau = \beta/2$ where both red and black expressions coincide.

The expressions for the two Green's functions, however sharing similar structures, in the end involve different functions and *a priori* any simple relation between $G_{\text{in}}(x,t,\Lambda)$ and $G_{\text{ej}}(x,t,\Lambda)$ is unexpected. It is clear from the formula for $G_{\text{in}}(x,t,\Lambda)$ that it depends on the spin rapidity $\Lambda$ in a non-trivial way which is a sufficient condition for breaking of the equivalence between GCE and MCE.

*Refined KMS relation and the Riemann-Hilbert problem:* We now sketch a derivation of the refined detailed balance relation [62]. The idea of the proof relies on interpreting the two Green's function as solutions to the Riemann-Hilbert problem (RHP). The RHP in its simplest formulation is a problem of determining a function which is analytic everywhere but along the real line and which asymptotically approaches 1. The non-analytic behaviour along the real line is characterized by the jump condition: $\chi_+(x) = \chi_-(x)\mathcal{G}(x)$, where $\chi_\pm(x) = \lim_{\epsilon \to 0} \chi(x \pm i\epsilon)$. For our application we need a correspondent generalization to matrix-valued functions [63, 64].

Representing the Green's functions through a solution to the RHP problem corresponds to identifying the jump matrices $\mathcal{G}_{\text{in,ej}}$. We then perform a series of manipulations of jump matrices which demonstrates that they can be made equal after an appropriate change of the coordinate and time. This allows us to conclude that

$$\frac{G_{\text{ej}}(-x,-t-i\beta/2)}{G_{\text{in}}(x,t-i\beta/2)} = e^{\beta\mu}W(\beta,\Lambda), \qquad (29)$$

with a *space-time-independent* function

$$W(\beta,\Lambda) = \frac{\det(1 + V_{\text{ej}}(-x,-t-i\beta/2))}{\det(1 + V_{\text{in}}(x,t-i\beta/2))}. \qquad (30)$$

Therefore $W(\beta,\Lambda)$ can be found by considering $t = 0$ and taking $x \to \infty$ limit, where the $x$-asymptote of the Fredholm determinants can be evaluated with the effective form-factors approach [65–67]. The result is $W(\beta,\Lambda) = \exp(\beta\mathcal{F}(\Lambda))$, thus finishing the proof of the refined KMS relation. [68]

The Green's function can be evaluated numerically [69]. This requires a numerical evaluation of the Fredholm determinant and can be done by the quadrature method [70]. In Fig. 2 we show the KMS relation between the *injection* and *ejection* Green's function. This constitutes a numerical proof of this relation.

*Non-interacting limit.* In the limit of vanishing impurity-gas interactions, the Green's functions become those of a free particle with its dispersion controlled by $\Lambda$ [71] and additionally for the ejection weighted by the thermal gas distribution

$$G_{\text{in}}(x,t,\Lambda) = e^{ix\Lambda - it\Lambda^2/2}, \qquad (31)$$

$$G_{\text{ej}}(x,t,\Lambda) = \sigma(\Lambda)e^{-ix\Lambda + it\Lambda^2/2}. \qquad (32)$$

In the momentum space the correlators are then proportional to $\delta(k \pm \Lambda)$ — in the non-interacting limit the spin rapidity $\Lambda$ can be identified with momentum $k$. This shows that the $\Lambda$ degree of freedom emerges from the interacting nature of the system. It is an open question how to observe this effect for non-integrable models, for example using the variational method [47, 48].

*Conclusions:* In this work we studied exact thermal Green's functions of an interacting impurity model. We have shown that due to the interactions the theory acquires a new thermodynamic parameter, the spin rapidity $\Lambda$ of the nested Bethe Ansatz. As a result, the ejection Green's function involves a sum over an ensemble of systems weighted with $\Lambda$-dependent free energies. The injection Green's function can also be resolved into contributions from different $\Lambda$'s. This appears as a result of quantum-mechanical averaging rather then thermal av-eraging. Despite that, the two Λ-resolved Green's functions obey the KMS relation. The existence of such a relation effectively promotes Λ to a thermodynamic parameter describing the equilibrium state of the system. On the other hand the KMS relation is a statement about analytic continuation of correlation functions. In order to prove it, it requires non-perturbative control over the correlators, which is provided within our mathematical framework. We then employed the Riemann-Hilbert problem to prove the KMS relation and we further verified it numerically.

The methods developed here apply beyond the thermal equilibrium. For example, the expressions for the Λ-resolved correlation functions are valid for *any* distribution $\sigma(k)$, not necessarily thermal. An interesting question in this direction is about the existence of KMS relation beyond the thermal equilibrium. Closely related statements of detailed balance were found in similar circumstances for the 1d interacting Bose gas [72, 73].

*Acknowledgments:* OG acknowledges the support from the Polish National Agency for Academic Exchange (NAWA) through the Grant No. PPN/ULM/2020/1/00247. MP and FTS acknowledge the support from the National Science Centre, Poland, under the SONATA grant 2018/31/D/ST3/03588 and POLONEZ BIS grant No. 2021/43/P/ST2/02904, respectively. This research is part of the project No. 2021/43/P/ST2/02904 co-funded by the National Science Centre and the European Union Framework Programme for Research and Innovation Horizon 2020 under the Marie Skłodowska-Curie grant agreement No. 945339. For the purpose of Open Access, the authors have applied a CC-BY public copyright licence to any Author Accepted Manuscript (AAM) version arising from this submission.

---

* Correspondence to: og@lims.ac.uk
[1] R. Kubo, Statistical-mechanical theory of irreversible processes. i. general theory and simple applications to magnetic and conduction problems, Journal of the Physical Society of Japan **12**, 570 (1957), https://doi.org/10.1143/JPSJ.12.570.
[2] P. C. Martin and J. Schwinger, Theory of many-particle systems. i, Phys. Rev. **115**, 1342 (1959).
[3] R. Haag, N. M. Hugenholtz, and M. Winnink, On the equilibrium states in quantum statistical mechanics, Communications in Mathematical Physics **5**, 215 (1967).
[4] W. Pusz and S. L. Woronowicz, Passive states and kms states for general quantum systems, Communications in Mathematical Physics **58**, 273 (1978).
[5] O. Bratteli and D. W. Robinson, *Operator Algebras and Quantum Statistical Mechanics 2* (Springer, 2002).
[6] L. Onsager, Reciprocal relations in irreversible processes. ii., Phys. Rev. **38**, 2265 (1931).
[7] N. G. van Kampen, *Stochastic Processes in Physics and Chemistry* (Elsevier, 2007).
[8] C. Jarzynski, Equalities and inequalities: Irreversibility and the second law of thermodynamics at the nanoscale, Annual Review of Condensed Matter Physics **2**, 329 (2011), https://doi.org/10.1146/annurev-conmatphys-062910-140506.
[9] F. M. Haehl, R. Loganayagam, P. Narayan, A. A. Nizami, and M. Rangamani, Thermal out-of-time-order correlators, KMS relations, and spectral functions, JHEP **12**, 154, arXiv:1706.08956 [hep-th].
[10] M. B. Zvonarev, V. V. Cheianov, and T. Giamarchi, Spin dynamics in a one-dimensional ferromagnetic bose gas, Phys. Rev. Lett. **99**, 240404 (2007).
[11] M. B. Zvonarev, V. V. Cheianov, and T. Giamarchi, Dynamical properties of the one-dimensional spin-1/2 bose-hubbard model near a mott-insulator to ferromagnetic-liquid transition, Phys. Rev. Lett. **103**, 110401 (2009).
[12] D. M. Gangardt and A. Kamenev, Bloch oscillations in a one-dimensional spinor gas, Phys. Rev. Lett. **102**, 070402 (2009).
[13] A. Lamacraft, Dispersion relation and spectral function of an impurity in a one-dimensional quantum liquid, Phys. Rev. B **79**, 241105 (2009).
[14] F. Massel, A. Kantian, A. J. Daley, T. Giamarchi, and P. Törmä, Dynamics of an impurity in a one-dimensional lattice, New Journal of Physics **15**, 045018 (2013).
[15] A. Kantian, U. Schollwöck, and T. Giamarchi, Competing regimes of motion of 1d mobile impurities, Physical Review Letters **113**, 10.1103/physrevlett.113.070601 (2014).
[16] M. Schecter, D. M. Gangardt, and A. Kamenev, Dynamics and Bloch oscillations of mobile impurities in one-dimensional quantum liquids, Annals of Physics **327**, 639 (2012).
[17] M. Schecter, A. Kamenev, D. M. Gangardt, and A. Lamacraft, Critical velocity of a mobile impurity in one-dimensional quantum liquids, Phys. Rev. Lett. **108**, 207001 (2012).
[18] E. Burovski, V. Cheianov, O. Gamayun, and O. Lychkovskiy, Momentum relaxation of a mobile impurity in a one-dimensional quantum gas, Physical Review A **89**, 10.1103/physreva.89.041601 (2014).
[19] O. Gamayun, O. Lychkovskiy, and V. Cheianov, Kinetic theory for a mobile impurity in a degenerate tonks-girardeau gas, Phys. Rev. E **90**, 032132 (2014).
[20] O. Gamayun, Quantum boltzmann equation for a mobile impurity in a degenerate tonks-girardeau gas, Phys. Rev. A **89**, 063627 (2014).
[21] O. Gamayun, O. Lychkovskiy, E. Burovski, M. Malcomson, V. V. Cheianov, and M. B. Zvonarev, Impact of the injection protocol on an impurity's stationary state, Phys. Rev. Lett. **120**, 220605 (2018).
[22] R. S. Christensen, J. Levinsen, and G. M. Bruun, Quasiparticle properties of a mobile impurity in a bose-einstein condensate, Phys. Rev. Lett. **115**, 160401 (2015).
[23] F. Grusdt, All-coupling theory for the fröhlich polaron, Physical Review B **93**, 10.1103/physrevb.93.144302 (2016).
[24] Y. E. Shchadilova, R. Schmidt, F. Grusdt, and E. Demler, Quantum dynamics of ultracold bose polarons, Phys. Rev. Lett. **117**, 113002 (2016).
[25] F. Grusdt, G. E. Astrakharchik, and E. Demler, Bose polarons in ultracold atoms in one dimension: beyond the fröhlich paradigm, New Journal of Physics **19**, 103035

[26] L. A. P. Ardila, Monte carlo methods for impurity physics in ultracold bose quantum gases, Nature Reviews Physics **4**, 214 (2022).

[27] S. Palzer, C. Zipkes, C. Sias, and M. Köhl, Quantum Transport through a Tonks-Girardeau Gas, Phys. Rev. Lett. **103**, 150601 (2009).

[28] J. Catani, G. Lamporesi, D. Naik, M. Gring, M. Inguscio, F. Minardi, A. Kantian, and T. Giamarchi, Quantum dynamics of impurities in a one-dimensional bose gas, Phys. Rev. A **85**, 023623 (2012).

[29] T. Fukuhara, A. Kantian, M. Endres, M. Cheneau, P. Schauß, S. Hild, D. Bellem, U. Schollwöck, T. Giamarchi, C. Gross, I. Bloch, and S. Kuhr, Quantum dynamics of a mobile spin impurity, Nature Physics **9**, 235 (2013).

[30] F. Meinert, M. Knap, E. Kirilov, K. Jag-Lauber, M. B. Zvonarev, E. Demler, and H.-C. Nägerl, Bloch oscillations in the absence of a lattice, Science **356**, 945 (2017).

[31] M. Serbyn, D. A. Abanin, and Z. Papić, Quantum many-body scars and weak breaking of ergodicity, Nature Physics **17**, 675 (2021).

[32] S. Moudgalya, B. A. Bernevig, and N. Regnault, Quantum many-body scars and hilbert space fragmentation: a review of exact results, Reports on Progress in Physics **85**, 086501 (2022).

[33] J. M. Deutsch, Quantum statistical mechanics in a closed system, Phys. Rev. A **43**, 2046 (1991).

[34] M. Srednicki, Chaos and quantum thermalization, Phys. Rev. E **50**, 888 (1994), cond-mat/9403051.

[35] M. Rigol, V. Dunjko, and M. Olshanii, Thermalization and its mechanism for generic isolated quantum systems, Nature **452**, 854 (2008).

[36] A. M. Kaufman, M. E. Tai, A. Lukin, M. Rispoli, R. Schittko, P. M. Preiss, and M. Greiner, Quantum thermalization through entanglement in an isolated many-body system, Science **353**, 794 (2016), https://www.science.org/doi/pdf/10.1126/science.aaf6725.

[37] C. Neill, P. Roushan, M. Fang, Y. Chen, M. Kolodrubetz, Z. Chen, A. Megrant, R. Barends, B. Campbell, B. Chiaro, A. Dunsworth, E. Jeffrey, J. Kelly, J. Mutus, P. J. J. O'Malley, C. Quintana, D. Sank, A. Vainsencher, J. Wenner, T. C. White, A. Polkovnikov, and J. M. Martinis, Ergodic dynamics and thermalization in an isolated quantum system, Nature Physics **12**, 1037 (2016).

[38] J. M. Deutsch, Eigenstate thermalization hypothesis, Reports on Progress in Physics **81**, 082001 (2018).

[39] S. Sugimoto, R. Hamazaki, and M. Ueda, Test of the eigenstate thermalization hypothesis based on local random matrix theory, Phys. Rev. Lett. **126**, 120602 (2021).

[40] C. N. Yang, Some exact results for the many-body problem in one dimension with repulsive delta-function interaction, Phys. Rev. Lett. **19**, 1312 (1967).

[41] The KMS relation for normalized Green's function involves the ratio of the partition functions. This is an artifact of working with a fixed number of impurities and can be avoided by considering the grand canonical ensemble also for them. The ratio of the partition functions leads then to the usual $e^{\beta \mu_{\rm imp}}$ term with $\mu_{\rm imp}$ the chemical potential for the impurity particles.

[42] A. Schirotzek, C.-H. Wu, A. Sommer, and M. W. Zwierlein, Observation of fermi polarons in a tunable fermi liquid of ultracold atoms, Phys. Rev. Lett. **102**, 230402 (2009).

[43] C. Kohstall, M. Zaccanti, M. Jag, A. Trenkwalder, P. Massignan, G. M. Bruun, F. Schreck, and R. Grimm, Metastability and coherence of repulsive polarons in a strongly interacting fermi mixture, Nature **485**, 615 (2012).

[44] M. Cetina, M. Jag, R. S. Lous, J. T. M. Walraven, R. Grimm, R. S. Christensen, and G. M. Bruun, Decoherence of impurities in a fermi sea of ultracold atoms, Phys. Rev. Lett. **115**, 135302 (2015).

[45] M. Cetina, M. Jag, R. S. Lous, I. Fritsche, J. T. M. Walraven, R. Grimm, J. Levinsen, M. M. Parish, R. Schmidt, M. Knap, and E. Demler, Ultrafast many-body interferometry of impurities coupled to a fermi sea, Science **354**, 96 (2016).

[46] Z. Yan, P. B. Patel, B. Mukherjee, R. J. Fletcher, J. Struck, and M. W. Zwierlein, Boiling a unitary fermi liquid, Phys. Rev. Lett. **122**, 093401 (2019).

[47] W. E. Liu, Z.-Y. Shi, J. Levinsen, and M. M. Parish, Radio-frequency response and contact of impurities in a quantum gas, Phys. Rev. Lett. **125**, 065301 (2020).

[48] W. E. Liu, Z.-Y. Shi, M. M. Parish, and J. Levinsen, Theory of radio-frequency spectroscopy of impurities in quantum gases, Phys. Rev. A **102**, 023304 (2020).

[49] A. C. Cassidy, C. W. Clark, and M. Rigol, Generalized thermalization in an integrable lattice system, Phys. Rev. Lett. **106**, 140405 (2011).

[50] Q.-Q. Wang, S.-J. Tao, W.-W. Pan, Z. Chen, G. Chen, K. Sun, J.-S. Xu, X.-Y. Xu, Y.-J. Han, C.-F. Li, and G.-C. Guo, Experimental verification of generalized eigenstate thermalization hypothesis in an integrable system, Light: Science & Applications **11**, 194 (2022).

[51] Please see Section S1 of the Supplementary Material for details of the Bethe ansatz solution.

[52] J. B. McGuire, Interacting fermions in one dimension. i. repulsive potential, Journal of Mathematical Physics **6**, 432 (1965).

[53] J. B. McGuire, Interacting fermions in one dimension. II. attractive potential, Journal of Mathematical Physics **7**, 123 (1966).

[54] H. Castella, X. Zotos, and P. Prelovšek, Integrability and ideal conductance at finite temperatures, Phys. Rev. Lett. **74**, 972 (1995).

[55] O. Gamayun, M. Panfil, and F. T. Sant'Ana, Mobile impurity in a one-dimensional gas at finite temperatures, Phys. Rev. A **106**, 023305 (2022).

[56] C. Recher and H. Kohler, From hardcore bosons to free fermions with painlevé v, Journal of Statistical Physics **147**, 542 (2012).

[57] O. Gamayun, A. G. Pronko, and M. B. Zvonarev, Impurity green's function of a one-dimensional fermi gas, Nuclear Physics B **892**, 83 (2015).

[58] O. Gamayun, A. G. Pronko, and M. B. Zvonarev, Time and temperature-dependent correlation function of an impurity in one-dimensional fermi and tonks–girardeau gases as a fredholm determinant, New Journal of Physics **18**, 045005 (2016).

[59] O. Gamayun, O. Lychkovskiy, and M. B. Zvonarev, Zero temperature momentum distribution of an impurity in a polaron state of one-dimensional Fermi and Tonks-Girardeau gases, SciPost Phys. **8**, 53 (2020).

[60] Please see Section S2 of the Supplemental Materials.

[61] Please see Section S4 of the Supplementary Material for more information on Fredholm determinants.

[62] Please see Section S3 of the Supplementary Material for

the formulation of the corresponding Riemann-Hilbert problem.
[63] K. McLaughlin and X. Zhou, *Recent Developments in Integrable Systems and Riemann-Hilbert Problems: AMS Special Session, Integrable Systems and Riemann-Hilbert Problems, November 10-12, 2000, University of Alabama, Birmingham, Alabama*, Contemporary mathematics - American Mathematical Society (American Mathematical Society, 2003).
[64] T. Bothner, On the origins of Riemann–Hilbert problems in mathematics*, Nonlinearity **34**, R1 (2021).
[65] O. Gamayun, N. Iorgov, and Y. Zhuravlev, Effective free-fermionic form factors and the XY spin chain, SciPost Phys. **10**, 70 (2021).
[66] Y. Zhuravlev, E. Naichuk, N. Iorgov, and O. Gamayun, Large time and long distance asymptotics of the thermal correlators of the impenetrable anyonic lattice gas (2021), arXiv:2110.06860 [cond-mat.quant-gas].
[67] D. Chernowitz and O. Gamayun, On the dynamics of free-fermionic tau-functions at finite temperature (2022), arXiv:2110.08194 [cond-mat.stat-mech].
[68] For more details please see section S3 in the Supplementary Materials.
[69] Please see Section S4 of the Supplementary Material for the details on numerical evaluation of the Green's functions and additional results.
[70] F. Bornemann, On the numerical evaluation of Fredholm determinants, Mathematics of Computation **79**, 871 (2009).
[71] In order to perform the limit, we shift $\Lambda \to \alpha\Lambda$ and then send $\alpha \to 0$.
[72] J. D. Nardis and M. Panfil, Exact correlations in the Lieb-Liniger model and detailed balance out-of-equilibrium, SciPost Phys. **1**, 015 (2016).
[73] J. De Nardis, M. Panfil, A. Gambassi, L. Cugliandolo, R. Konik, and L. Foini, Probing non-thermal density fluctuations in the one-dimensional bose gas, SciPost Physics **3**, 10.21468/scipostphys.3.3.023 (2017).

# Supplemental Material
# Kubo-Martin-Schwinger relation for an interacting mobile impurity

## Contents

## S1 Yang-Gaudin model and the impurity problem

In this section we review the Bethe ansatz solution of the Yang-Gaudin model. We focus on the sectors of the theory with zero or one spin up particle in a thermodynamically large sea of spin down particles. These sectors are relevant for the impurity Green's function. In our presentation we follow [1, 2]. We also discuss the excitation spectrum of the theory, specifically we show a degeneracy excitations over finite temperature state. This degeneracy can be traced back as a microscopic foundation for the refined detailed balance.

The Hamiltonian of the Yang-Gaudin model is

$$H = \int_0^L \mathrm{d}x \left( -\frac{\hbar^2}{2m} \sum_{\sigma=\uparrow,\downarrow} \psi_\sigma^\dagger(x) \partial_x^2 \psi_\sigma(x) + g \psi_\uparrow^\dagger(x)\psi_\uparrow(x)\psi_\downarrow^\dagger(x)\psi_\downarrow(x) \right). \qquad (\mathrm{S1.1})$$

The Hamiltonian commutes with the number operators of spin up and down particles and with the total momentum operator,

$$N_\sigma = \int \mathrm{d}x\, \psi_\sigma^\dagger(x)\psi_\sigma(x), \qquad (\mathrm{S1.2})$$

$$P = -i\hbar \int \mathrm{d}x \sum_{\sigma=\uparrow,\downarrow} \psi_\sigma^\dagger(x)\partial_x \psi_\sigma(x) \qquad (\mathrm{S1.3})$$

and particle number of each type is a conserved quantity. The eigenstates of the system can be then characterized by total number of particles $N = N_\uparrow + N_\downarrow$ and $M = N_\downarrow$ number of spin down particles. We call tuple $(N, M)$ a sector. In each sector $(N, M)$ the Hamiltonian can be written in the first quantized form

$$H_{N,M} = -\frac{\hbar^2}{2m} \sum_{j=1}^{N-M} \frac{\partial^2}{\partial x_{\uparrow,j}^2} - \frac{\hbar^2}{2m} \sum_{j=1}^{M} \frac{\partial^2}{\partial x_{\downarrow,j}^2} + g \sum_{j=1}^{N-M} \sum_{k=1}^{M} \delta(x_{\uparrow,j} - x_{\downarrow,k}). \qquad (\mathrm{S1.4})$$

In the following we set $\hbar = 1$ and $m = 1$.

**The Bethe ansatz solution**

The eigenstates in the sector $(N, M)$ are characterized by two sets of rapidities (quasi-momenta), $\{k\} = k_1, \ldots, k_N$ and $\{\lambda\} = \lambda_1, \ldots, \lambda_M$ which are real numbers and obey Pauli principle in each set separately. The basis of the Fock space of the model is constructed in the standard way by defining a vacuum state $|0\rangle$ such that

$$\psi_\sigma |0\rangle = 0, \qquad \langle 0| \psi_\sigma^\dagger = 0, \qquad \langle 0|0\rangle = 1. \qquad (\mathrm{S1.5})$$

The eigenstates take then the following form

$$|\Psi_{N,M}(\{k\},\{\lambda\})\rangle = \int d^N x \sum_{\{s\}} \Psi_{N,M}^{\{s\}}(\{k\},\{\lambda\}|\{x\}) \psi_{s_1}^\dagger(x_1)\ldots\psi_{s_N}^\dagger(x_N)|0\rangle, \quad \text{(S1.6)}$$

with wave function $\Psi_{N,M}^{\{s\}}(\{k\},\{\lambda\}|\{x\})$ and set of positions $\{x\} = x_1,\ldots,x_N$. For the wave function in sector $(N,M)$ to be non-zero the set of spins $\{\sigma\} = \sigma_1,\ldots,\sigma_N$ must be such that exactly $M$ elements among them are spin downs. Finally, the rapidities are solutions to the nested Bethe equations

$$e^{ik_j L} = \prod_{a=1}^M \frac{k_j - \lambda_a + ic/2}{k_j - \lambda_a - ic/2}, \quad j = 1,\ldots,N, \quad \text{(S1.7)}$$

$$\prod_{j=1}^N \frac{k_j - \lambda_a + ic/2}{k_j - \lambda_a - ic/2} = \prod_{b=1}^M \frac{\lambda_b - \lambda_a + ic}{\lambda_b - \lambda_a - ic}, \quad a = 1,\ldots,M. \quad \text{(S1.8)}$$

The momentum and energy of an eigenstate are

$$P(\{k\},\{\lambda\}) = \sum_{j=1}^N k_j, \quad E(\{k\},\{\lambda\}) = \frac{1}{2}\sum_{j=1}^N k_j^2. \quad \text{(S1.9)}$$

The wave function $\Psi_{N,M}^{\{s\}}(\{k\},\{\lambda\}|\{x\})$ takes the typical structure of Bethe ansatz solvable models, it consists of a superposition of plane waves with rapidity-dependent amplitudes. In the following we focus on sectors $(N,0)$ and $(N+1,1)$ which are relevant for the impurity problem.

In sector $(N,0)$ the system is a free Fermi gas of $N$ spin up particles. The second set of rapidities is empty, $\{\lambda\} = 0$ and the Bethe equations reduce to the standard quantization conditions

$$k_j = \frac{2\pi}{L} I_j, \quad I_j \in \mathbb{Z}, \quad \text{(S1.10)}$$

with the wave function given by a Slater determinant of an $N \times N$ matrix,

$$\Psi_{N,0}^{\uparrow,\ldots,\uparrow}(\{k\},\{\lambda\}|\{x\}) = \frac{1}{N!L^{N/2}} \det\left(e^{iq_j x_k}\right). \quad \text{(S1.11)}$$

In sector $(N+1,1)$ the wave-function can also be represented by a determinant. To achieve this one changes the coordinate system to the impurity's (the spin down particle) rest frame, a transformation known as the Lee-Low-Pines transformation [3], applied to the McGuire model in [4, 5]. Here we follow [2]. The resulting wave-function is

$$\Phi(y_1,\ldots,y_N) = \frac{C_\Phi}{N!L^{N/2}} \det\begin{pmatrix} e^{ik_1 x_1} & \cdots & e^{ik_{N+1} x_1} \\ \vdots & \ddots & \vdots \\ e^{ik_1 x_N} & \cdots & e^{ik_{N+1} x_N} \\ \nu_-(k_1) & \cdots & \nu_-(k_{N+1}) \end{pmatrix}, \quad \text{(S1.12)}$$

with the normalization factor

$$C_\Phi^{-2} = \prod_{j=1}^{N+1}\left(1 + \frac{4}{gL}\nu_-(k_j)\nu_+(k_j)\right) \sum_{j=1}^{N+1} \frac{\nu_-(k_j)\nu_+(k_j)}{1 + \frac{4}{gL}\nu_-(k_j)\nu_+(k_j)}, \quad \text{(S1.13)}$$

and with

$$\nu_\pm(q) = \frac{g}{2}\frac{1}{q - k_{mp}}, \quad k_\pm = \frac{g}{2}(\Lambda \pm i), \quad \text{(S1.14)}$$

and is related to the original wave function through two relations

$$\tilde\Psi(\{x\}) = e^{iPx_{N+1}}\Phi_N(x_1 - x_{N+1},\ldots,x_N - x_{N+1}), \quad \text{(S1.15)}$$

$$\tilde\Psi(\{x\}) = \sum_{j=1}^{N+1}(-1)^{N-j}\Psi^{\downarrow_j}(x_1,\ldots,x_{j-1},x_{N+1},x_{j+1},\ldots,x_N), \quad \text{(S1.16)}$$

where by $\downarrow_j$ we abbreviate a set of spins $\{\sigma\}$ with a single spin down at position $j$ in the sea of spin up particles. The rapidities $\{k_j\}$ and the single spin rapidity $\Lambda$ obey

$$e^{ik_l L} = \frac{k_l - \Lambda + ic/2}{k_l - \Lambda - ic/2}, \qquad l = 1, \ldots, N+1, \tag{S1.17}$$

$$1 = \prod_{l=1}^{N+1} \frac{k_l - \Lambda + ic/2}{k_l - \Lambda - ic/2}. \tag{S1.18}$$

The second equation can be understood as the quantization condition of the total momentum. Indeed taking the product over all $l$ of the first equation and then using the second equation we find

$$e^{iPL} = \prod_{l=1}^{N+1} \frac{k_l - \Lambda + ic/2}{k_l - \Lambda - ic/2} = 1, \tag{S1.19}$$

**Bethe equations and excited states**

The Bethe equations in the logarithmic form are

$$k_j = \frac{2\pi}{L}\left(n_j - \frac{\delta(k_j)}{\pi}\right), \quad j = 1, \ldots, N+1 \tag{S1.20}$$

$$\sum_{j=1}^{N+1} k_j = \frac{2\pi n}{L}. \tag{S1.21}$$

where quantum numbers $n_j$ are integers and obey the Pauli principle. The phase shift is

$$\delta(k) = \frac{\pi}{2} - \arctan\left(\Lambda - \alpha k\right), \qquad \alpha = \frac{2\pi}{\gamma}. \tag{S1.22}$$

The spin rapidity $\Lambda$ can be fixed by specifying values of other integrals of motions in this model. Traditionally, we require that the total momentum given by

$$P(\{k_j\}, \Lambda) = \sum_{k=1}^{N+1} k_j, \tag{S1.23}$$

is fixed, i.e. $P(\{k_j\}, \Lambda) = Q$. The $\Lambda$ dependence in (S1.23) is implicit through $k_j$ as solutions to the Bethe equations. Therefore, for given $Q$ and the set of integers one can resolve condition (S1.23) and thus solve the Bethe equations.

The Bethe equations for rapidities $\{k_j\}$ can be rewritten in terms of a single function obeying a transcendental equation,

$$k_j = \frac{2\pi}{L}\left[n_j - \frac{1}{2} + \frac{1}{\pi}\arctan f\left(\Lambda - \frac{2\pi\alpha}{L}n_j\right)\right], \tag{S1.24}$$

where

$$f(x) = x + \frac{2\pi\alpha}{L}\left(\frac{1}{2} - \frac{1}{\pi}\arctan f(x)\right). \tag{S1.25}$$

Function $f$ depends only on the parameters of the system $\alpha$ and $L$ but not on the rapidities. The expression for $k_j$ gives a bound on the total momentum of the state. Denoting

$$K = \frac{2\pi}{L}\sum_{j=1}^{N+1} n_j, \tag{S1.26}$$

and using that $\arctan(x)$ is a bounded function we find

$$K(\{n_j\}) - \frac{2\pi(N+1)}{L} \leq Q \leq K(\{n_j\}). \tag{S1.27}$$

The maximal and minimal values of rapidity correspond to $\Lambda = \pm\infty$. In that case the rapidities can be computed analytically. The solution is

$$k_j^+ = \frac{2\pi}{L} n_j, \qquad \Lambda = \infty, \tag{S1.28}$$

$$k_j^- = \frac{2\pi}{L} (n_j - 1), \qquad \Lambda = -\infty. \tag{S1.29}$$

For the other cases $\Lambda$ is finite and equations have to be solved numerically.

**Degeneration of states:** Consider a set of quantum numbers $\{n_j\}$ and corresponding $\Lambda$ such that the Bethe equations are fulfilled and denote rapidities by $k_j$. We can now build another state related to it by a parity operation. Choose $\bar{n}_j = -n_j + 1$ and $\bar{\Lambda} = -\Lambda$. The corresponding rapidities are $\bar{k}_j = -k_j$ and obey the Bethe equations. The two states have the same energy and opposite momenta.

Derivation

$$\begin{aligned}\bar{k}_j = -k_j &= -\frac{2\pi}{L} \left[ n_j - \frac{1}{2} + \frac{1}{\pi} \arctan f \left( \Lambda - \frac{2\pi\alpha}{L} n_j \right) \right] \\ &= \frac{2\pi}{L} \left[ \bar{n}_j - \frac{1}{2} - \frac{1}{\pi} \arctan f \left( \Lambda + \frac{2\pi\alpha}{L} \bar{n}_j - \frac{2\pi\alpha}{L} \right) \right] \\ &= \frac{2\pi}{L} \left[ \bar{n}_j - \frac{1}{2} + \frac{1}{\pi} \arctan f \left( \bar{\Lambda} - \frac{2\pi\alpha}{L} \bar{n}_j \right) \right], \end{aligned} \tag{S1.30}$$

where in the last step we used the symmetry property of $f(x)$ and that $\arctan(x)$ is an odd function. This can be also derived directly from the original formulation.

**Excitations far from the ground state:** In the thermodynamically large system, a state can be described by the distribution of rapidities $\sigma(k)$ and value of the rapidity $\Lambda$. With respect to such state we consider excitations. This take form of modifying some subset of quantum numbers $n_j$ and/or adding/removing the impurity.

Consider a state without the impurity described by quantum numbers $\{n_j\}_{j=1}^N$ and a state with the impurity described by quantum numbers $\{n'_j\}_{j=1}^{N+1}$ and $\Lambda$. We assume that the two sets share most of the quantum numbers with $m$ quantum numbers different such that $m/L \to 0$ in the thermodynamic limit. Then the two states are described by the same distribution $n(k)$. Consider subleading in the system size corrections to the values of the rapidities. For $j = 1, \ldots, N$,

$$k'_j - k_j = \frac{2\pi}{L} \left[ n'_j - n_j + \frac{1}{\pi} \arctan f \left( \Lambda - \frac{2\pi\alpha}{L} n'_j \right) \right]. \tag{S1.31}$$

Unless $n'_j - n_j \sim L$, the difference between the rapidities is of the order $1/L$. For the quantum numbers that were not modified we then have

$$k'_j - k_j = \frac{2}{L} \arctan f(\Lambda - \alpha k_j), \tag{S1.32}$$

where we replaced the quantum number $2\pi n'_j/L$ by the rapidity $k_j$ as the error is of the order $1/L$ and therefore has subleading contribution to $k_j - k'_j$.

Consider now difference in momenta between the two states

$$\begin{aligned} p &= \sum_{j=1}^{N+1} k'_j - \sum_{j=1}^{N} k_j = k'_{N+1} + \sum_{j=1}^{m} (p_j - h_j) + \sum_{j=1}^{N} (k'_j - k_j) \\ &= \sum_{j=1}^{m+1} p_j - \sum_{j=1}^{m} h_j + \frac{1}{\pi} \int dk\, \sigma(k) \arctan f(\Lambda - \alpha k). \end{aligned} \tag{S1.33}$$

Similar computations for the energy difference give

$$\omega = \frac{1}{2} \sum_{j=1}^{m+1} p_j^2 - \frac{1}{2} \sum_{j=1}^{m} h_j^2 + \frac{1}{\pi} \int dk\, \sigma(k) k \arctan f(\Lambda - \alpha k). \tag{S1.34}$$

These excitation can be summarized by

$$n(k) \longrightarrow n(k), \{p_j\}_{j=1}^{m+1}, \{h_j\}_{j=1}^{m}, \Lambda \tag{S1.35}$$

We can also consider an opposite type of excitation which involves annihilating the impurity,

$$n(k), \Lambda' \longrightarrow n(k), \{p'_j\}_{j=1}^{m}, \{h'_j\}_{j=1}^{m+1}. \tag{S1.36}$$

The momentum and energy carried by this excitation is

$$p' = \sum_{j=1}^{m} p_j - \sum_{j=1}^{m+1} h_j - \frac{1}{\pi} \int dk\, \sigma(k) \arctan f\left(\Lambda' - \alpha k\right) \tag{S1.37}$$

$$\omega' = \frac{1}{2}\sum_{j=1}^{m} p_j^2 - \frac{1}{2}\sum_{j=1}^{m+1} h_j^2 - \frac{1}{\pi} \int dk\, \sigma(k) k \arctan f\left(\Lambda' - \alpha k\right). \tag{S1.38}$$

We now show that there exist a symmetry between the two types of excitations. This symmetry underlies the refined detailed balance. Choosing $\Lambda' = -\Lambda$, $p'_j = h_j$ and $h'_j = p_j$ the two excited states carry exactly opposite momenta and energies with respect to their background states if $n(k)$ is a symmetric function of $k$. Reformulating this, for every excited state in which we create an impurity there exist an excited state in which we annihilate the impurity such that the two excitations have exactly opposite energies, momenta and $\Lambda$'s.

## S2 Impurity Green's functions at finite temperatures

In this Section we derive the expression for the ejection Green's function at finite temperature and show that it breaks the Eigenstate Thermalization Hypothesis. We also derive explicit expression for the finite temperature ejection and injection Green's function. For the ejection Green's function we generalize the static finite temperature result of [6]. For the injection the finite tempearture dynamic Green's function was computed in [2].

We consider now evaluation of the ejection Green's function in a finite system. In thermal equilibrium it is given by

$$G_{\text{ej}}(x,t) = \text{tr}\left(\hat{\rho}\Psi^\dagger(x,t)\Psi(0,0)\right), \tag{S2.1}$$

with thermal density matrix

$$\hat{\rho} = \frac{1}{\mathcal{Z}} e^{-\beta(H-\mu N)}, \tag{S2.2}$$

Denoting eigenstates in finite system by $|\{k\},\Lambda\rangle$ we write

$$G_{\text{ej}}(x,t) = \frac{1}{\mathcal{Z}} \sum_{\{k\},\Lambda} e^{-\beta(E-\mu N)} \langle\{k\},\Lambda|\Psi^\dagger(x,t)\Psi(0,0)|\{k\},\Lambda\rangle, \tag{S2.3}$$

with the partition function $\mathcal{Z}$ having analogous representation

$$\mathcal{Z} = \sum_{\{k\},\Lambda} e^{-\beta(E-\mu N)}. \tag{S2.4}$$

The building block of the finite temperature correlation function is an expectation value in a single eigenstate, which we denote

$$\langle\{k\},\Lambda|\Psi^\dagger(x,t)\Psi(0,0)|\{k\},\Lambda\rangle = \sum_{\{q\}} e^{ix(P_q-P_k)-it(E_q-E_k)} |\langle\{q\}|\Psi(0)|\{k\},\Lambda\rangle|^2. \tag{S2.5}$$

The normalized form-factors $\langle\{q\}|\Psi(0)|\{k\},\Lambda\rangle$ were computed in [1] and take the following form

$$|\langle\{k\},\Lambda|\Psi(0)|\{q\}\rangle|^2 = \left(\frac{2}{L}\right)^N \frac{\prod_{j=1}^{N+1} \frac{\partial k_j}{\partial \Lambda}}{\sum_{j=1}^{N+1} \frac{\partial k_j}{\partial \Lambda}} \det_{1\le j,k\le N}\left(\frac{1}{q_j-k_l} - \frac{1}{q_j-k_{N+1}}\right)^2. \tag{S2.6}$$

Here the derivative of $k_j$ over $\Lambda$ is formal and explicitly equals to

$$\frac{\partial k_j}{\partial \Lambda} = \frac{2}{L}\left(1 + (\alpha k_j - \Lambda)^2 + \frac{2\alpha}{L}\right)^{-1} \tag{S2.7}$$

Notice that in the sum in the denominator, one can ignore $1/L$ corrections and obtain

$$a(\Lambda) \equiv \sum_{j=1}^{N+1} \frac{\partial k_j}{\partial \Lambda} = \int \frac{dk}{\pi} \frac{\sigma(k)}{1 + (\alpha \Lambda - k)^2} + O(1/L) \tag{S2.8}$$

The result of [6], here generalized to dynamic correlation function, is that in the thermodynamic limit the expectation value $\langle \{k\}, \Lambda | \Psi^\dagger(x,t) \Psi(0,0) | \{k\}, \Lambda \rangle$ depends on the underlying distribution of $\{k_j\}$ and on the spin rapidity $\Lambda$

$$\langle \sigma(k), \Lambda | \Psi^\dagger(x,t) \Psi(0,0) | \sigma(k), \Lambda \rangle = \lim_{\text{td}} \langle \{k\}, \Lambda | \Psi^\dagger(x,t) \Psi(0,0) | \{k\}, \Lambda \rangle. \tag{S2.9}$$

This implies that when performing the thermal averaging one can use the saddle-point argument to localize the sum over $\{k_j\}$ at configuration that maximizes the free energy. The free energy of the system takes the following form [6]

$$\mathcal{F}(\sigma(k), \Lambda) = L\mathcal{F}_{\text{th}}(\sigma(k)) + \mathcal{F}_0(\sigma(k), \Lambda) + \text{const} + \mathcal{O}(1/L), \tag{S2.10}$$

where $\mathcal{F}_{\text{th}}(\sigma(k))$ corresponds to the free fermions energy, thence not depending on the rapidity $\Lambda$. However, the subleading contribution $\mathcal{F}(\sigma(k), \Lambda)$ does depend on it. Therefore, the thermal sum in the numerator of the ejection Green's function evaluates in the thermodynamic limit to

$$\text{const} \times e^{-L\mathcal{F}_{\text{th}}(\sigma(k))} \times \sum_\Lambda e^{-\beta \mathcal{F}(n(k), \Lambda)} \langle \sigma(k), \Lambda | \Psi^\dagger(x,t) \Psi(0,0) | \sigma(k), \Lambda \rangle, \tag{S2.11}$$

with constant factors coming from the saddle-point evaluation. The same factors will appear in the evaluation of the partition function and therefore they cancel in the final expression for the Green's function. This takes then the following form

$$G_{\text{ej}}(x,t) = \frac{\sum_\Lambda e^{-\beta \mathcal{F}(\sigma(k),\Lambda)} \langle \sigma(k), \Lambda | \Psi^\dagger(x,t) \Psi(0,0) | \sigma(k), \Lambda \rangle}{\sum_\Lambda e^{-\beta \mathcal{F}(\sigma(k),\Lambda)}}, \tag{S2.12}$$

Rewriting the sum over $\Lambda$ through integrals we obtain

$$G_{\text{ej}}(x,t) = \frac{1}{\mathcal{Z}_{\text{ej}}} \int \frac{d\Lambda}{2\pi} e^{-\beta \mathcal{F}(\Lambda)} G_{\text{ej}}(x,t,\Lambda), \tag{S2.13}$$

where we defined

$$\mathcal{Z}_{\text{ej}} = \int \frac{d\Lambda}{2\pi} a(\Lambda) e^{-\beta \mathcal{F}(\Lambda)} \tag{S2.14}$$

with $a(\Lambda)$ defined in (S2.8). Finally, because the saddle point configuration comes from the extremum of the free energy of the non-interacting Fermi gas, the distribution $\sigma(k)$ takes the Fermi-Dirac form

$$\sigma(k) = \frac{1}{1 + e^{\beta(k^2/2 - \mu)}}. \tag{S2.15}$$

The final formula for $G_{\text{ej}}(x,t,\Lambda)$ can be obtained from generalization of the static Green's function. In [6] it was derived

$$G_{\text{ej}}(x; \sigma, \Lambda) = \det_\sigma\left(1 + \hat{V} - \hat{W}_-\right) - \det_\sigma\left(1 + \hat{V}\right), \tag{S2.16}$$

The kernels are

$$\hat{V}(q, q') = \frac{e_+(q)e_-(q') - e_-(q)e_+(q')}{q - q'}, \qquad \hat{W}_-(q, q') = -\frac{1}{\pi} e_-(q)e_-(q'), \tag{S2.17}$$

with functions $e_\pm(q)$ given by

$$e_+(q) = \frac{1}{\pi} e^{iqx/2 + i\delta(q)}, \quad e_-(q) = e^{-iqx/2} \sin\delta(q). \tag{S2.18}$$

We also note that $G_{\rm ej}(x, -\Lambda)$ is a complex function such that $G_{\rm ej}(x, -\Lambda) = G^*_{\rm ej}(x, \Lambda)$. Therefore the resulting one-body function $G_{\rm ej}(x)$ is a real function.

To include the time dependence we include the energy contribution to functions $e_\pm(q)$,

$$e_+(q) = \frac{1}{\pi} e^{iqx/2 - itq^2/4 + i\delta(q)} - \frac{e_-(q)}{\pi} \int_{-\infty}^{\infty} \frac{dk}{\pi} \frac{e^{ikx - itk^2/2}}{q - k - i0}, \quad e_-(q) = e^{-iqx/2 + itq^2/4} \sin\delta(q). \tag{S2.19}$$

Notice that for $x > 0$ and $t = 0$ the integral in $e_+(q)$ vanishes. In this way we obtain the formula for the ejection Green's function $G_{\rm ej}(x, t; \sigma, \Lambda)$ reported in the main text.

**Injection Green's function:** Similar analysis can be performed for the injection Green's function. We find there the ETH works directly at the level of the full correlator reducing the thermal averaging to a single expectation value. However, as first investigated in [2], it is possible to truncate the internal sum to states with fixed value of $\Lambda$ thus in practice realizing the projection operator $P_\Lambda$. The formula for the $\Lambda$-resolved correlation function was derived in [2]. The expressions are

$$G_{\rm in}(x, t, \Lambda) = \det_\sigma(1 + \hat{V} - \hat{W}_+) - H \det_\sigma(1 + \hat{V}), \tag{S2.20}$$

Here $\det_{[-Q,Q]}(1 + \hat{K})$ denotes a Fredholm determinant of the kernel $K(q, q')$ with a measure given by the Fermi-Dirac distribution $\sigma(k)$ with $Q$ the Fermi rapidity. The kernels appearing above are

$$V(q, q') = \frac{1}{\pi} \frac{e_+(q) e_-(q') - e_+(q') e_-(q)}{q - q'}, \quad W_+(q, q') = \frac{1}{\pi} e_+(q) e_+(q'), \tag{S2.21}$$

where $e_+(q) = e(q) e_-(q)$ and functions $\epsilon_-(q)$ and $\epsilon(q)$ given by

$$e_-(q) = e^{itq^2/4 - ixq/2}, \tag{S2.22}$$

$$e(q) = \frac{F(q + i0) - F(\kappa_+)}{\alpha q - \Lambda - i} - \frac{F(q - i0) - F(\kappa_-)}{\alpha q - \Lambda + i}, \tag{S2.23}$$

and additionally $H = 1 - F(\kappa_+)/\alpha + F(\kappa_-)/\alpha$. Here $\kappa_\pm = (\Lambda \pm i)/\alpha$ and

$$F(z) = \int \frac{dk}{2\pi i} \frac{e^{-itk^2/2 + ikx}}{k - z} \tag{S2.24}$$

This function can be expressed in terms of error function. For a real argument we specify how we deform a contour to pass over or under a real line by $\pm i0$ shifts.

## S3 Riemann-Hilbert problem and the KMS relation

In this section, we reformulate the computation of Fredholm determinants in terms of a Riemann-Hilbert problem (RHP). We perform a transformation on the RHP for the injection and ejection cases to derive the refined detailed balance and mostly follow Ref. [7]. Let us start by recalling the statement of the matrix Riemann-Hilbert problem.

Let $\Sigma$ be an oriented contour in a complex plane and let $\mathcal{G}$ be a matrix valued function defined on $\Sigma$. The task is to find a matrix valued function $\chi(z)$ which is holomorphic in the complement of $\Sigma$. Additionally, let us denote by $\chi_\pm(z)$ the value of $\chi(q)$ when it approaches $\Sigma$ from one of its two sides. We require $\chi(z)$ to fulfill two conditions

- the limiting values are related through the jump matrix $\mathcal{G}$, i.e. $\chi_+ = \chi_-(1 + 2\pi i \mathcal{G})$,
- the function $\chi(z)$ approaches the identity matrix when $|z| \to \infty$ in the complement of the contour $\Sigma$.

In our case, the $\Sigma$ contour will be simply the real line.

Our strategy to prove the refined the detailed balance is the following. We will reformulate the computation of the two Green's as Riemann-Hilbert problems. This amounts to specifying the contour $\Sigma$ (which in both cases is real line) and the jump matrix. We will then perform some transformations of the RHP to arrive at the same jump matrix. From this, assuming the uniqueness of the solution to the RHP, we derive the following relation between the two Green's functions

$$\frac{G_{\mathrm{ej}}(-x,-t-i\beta/2)}{G_{\mathrm{in}}(x,t-i\beta/2)} = e^{\beta\mu}\frac{\det(1+V_{\mathrm{ej}})}{\det(1+V_{\mathrm{in}})}. \tag{S3.1}$$

Importantly, from the RHP we can infer that function the ratio of the determinants is independent of $x$ and $t$. Therefore it can be evaluated by taking a suitable limit of $t=0$ and $x\to\infty$. Computing the ratio amounts then to extracting the asymptotic behavior of the two determinants. This can be achieved with the effective form-factors approach. The result is

$$\frac{\det(1+V_{\mathrm{ej}})}{\det(1+V_{\mathrm{in}})} = e^{\beta\mathcal{F}_0(\Lambda)}, \tag{S3.2}$$

thus proving the refined detailed balance relation.

We start with the injection case and rewrite the kernel $V_{\mathrm{in}}$ in the vector notations

$$V_{\mathrm{in}}(q,q') = \frac{e_+(q)e_-(q') - e_+(q')e_-(q)}{q-q'} = \frac{\langle E_{\mathrm{in}}(q)|E_{\mathrm{in}}(q')\rangle}{q-q'}, \tag{S3.3}$$

where we have introduced "bra" and "ket" vectors

$$|E_{\mathrm{in}}(q)\rangle = \begin{pmatrix} e_-(q) \\ -e_+(q) \end{pmatrix}, \qquad \langle E(q)| = (e_+(q), e_-(q)). \tag{S3.4}$$

In these notations, we can immediately recognize the kernel as a generalized sine-kernel [8]. Notice also that in the original kernels for both injection and ejection Eqs. (S2.17),(S2.21) we can replace factors responsible for time dynamics $e^{itq^2/2} \to e^{it\varepsilon(q)}$, with the shifted energy $\varepsilon(q) = q^2/2 - \mu$, which results in the rescalings of the components $e_\pm(q) \to e_\pm(q)e^{\mp i\mu t/2}$ and the total correlation functions

$$G_{\mathrm{in}} \to e^{-i\mu t}G_{\mathrm{in}}, \qquad G_{\mathrm{ej}} \to e^{i\mu t}G_{\mathrm{ej}}. \tag{S3.5}$$

Because of the special (integrable) form of the operator $\hat{V}$ its resolvent $\hat{R}$ defined via

$$(1+\hat{V}_{\mathrm{in}})(1-\hat{R}_{\mathrm{in}}) = 1 \tag{S3.6}$$

also have an integrable form

$$R_{\mathrm{in}}(q,q') = \frac{\langle F_{\mathrm{in}}(q)|F_{\mathrm{in}}(q')\rangle}{q-q'}. \tag{S3.7}$$

Here, similar to (S3.4), we have put

$$|F_{\mathrm{in}}(q)\rangle = \begin{pmatrix} f_-(q) \\ -f_+(q) \end{pmatrix}, \qquad \langle F_{\mathrm{in}}(q)| = (f_+(q), f_-(q)). \tag{S3.8}$$

The components satisfy linear integral equations

$$|E_{\mathrm{in}}(q)\rangle = |F_{\mathrm{in}}(q)\rangle(1+\hat{V}_{\mathrm{in}}), \qquad \langle E_{\mathrm{in}}(q)| = (1+\hat{V}_{\mathrm{in}})\langle F_{\mathrm{in}}(q)|, \tag{S3.9}$$

where we remind that the action of the operator is given by the convolution with its kernel.

This system of linear integral equations can be reformulated as a Riemann-Hilbert problem. To do so we formally introduce $2\times 2$ matrices

$$\chi_{\mathrm{in}}(q) = 1 - \int dk \frac{|F_{\mathrm{in}}(k)\rangle\langle E_{\mathrm{in}}(k)|}{k-q}, \qquad \tilde{\chi}_{\mathrm{in}}(q) = 1 + \int dk \frac{|E_{\mathrm{in}}(k)\rangle\langle F_{\mathrm{in}}(k)|}{k-q}, \tag{S3.10}$$

and verify that

$$|F_{\mathrm{in}}(q)\rangle = \chi_{\mathrm{in}}(q)|E_{\mathrm{in}}(q)\rangle, \qquad \langle F_{\mathrm{in}}(q)| = \langle E_{\mathrm{in}}(q)|\tilde{\chi}_{\mathrm{in}}(q), \qquad \chi_{\mathrm{in}}(q)\tilde{\chi}_{\mathrm{in}}(q) = 1. \tag{S3.11}$$

The matrix function $\chi$ is analytic everywhere except the real line, where it experiences a jump with

$$\mathcal{G}_{\text{in}}(q) = |E_{\text{in}}(q)\rangle\langle E_{\text{in}}(q)| = \begin{pmatrix} e_-(q)e_+(q) & -(e_+(q))^2 \\ (e_-(q))^2 & -e_-(q)e_+(q), \end{pmatrix} \tag{S3.12}$$

Therefore $\chi_{\text{in}}(q)$ is a solution to the Riemann-Hilbert problem.

Formulation of $\chi(q)$ as a solution to the RHP allows us to infer about its asymptotics. Asymptotically, the solution to the RHP takes the following form

$$\chi_{\text{in}}(q) = 1 + \frac{1}{q}\begin{pmatrix} B_{-+} & B_{--} \\ -B_{++} & -B_{+-} \end{pmatrix} + \frac{1}{q^2}\begin{pmatrix} C_{-+} & C_{--} \\ -C_{++} & -C_{+-} \end{pmatrix} + o(q^{-2}), \tag{S3.13}$$

where using notations (S3.4) and (S3.8) we explicitly write

$$B_{\alpha\beta} = \int dk f_\alpha(k) e_\beta(k), \qquad C_{\alpha\beta} = \int dk f_\alpha(k) e_\beta(k) k. \tag{S3.14}$$

These expressions are also usually referred to as potentials. The symmetry of the kernel $V(q,q') = V(q',q)$ is reflected in the relation $B_{+-} = B_{-+}$. One can now easily express Green's function via potentials [7]. In particular, using relation (S3.9), for the injection Green's function we obtain

$$e^{it\mu}G_{\text{in}}(x,t,\Lambda) \equiv (h-1)\det(1+\hat{V}_{\text{in}}) + \det(1+\hat{V}_{\text{in}} - \hat{W}_+) = (h - B_{++})\tau_{\text{in}}, \tag{S3.15}$$

where we used that $W(q,q') = e_+(q)e_+(q')$ is rank one and introduced the $\tau$-function as the determinant of the operator $\hat{V}_{\text{in}}$,

$$\tau_{\text{in}} \equiv \det(1+\hat{V}_{\text{in}}). \tag{S3.16}$$

Now let us consider different RH problem for the matrix $\phi_{\text{in}}$

$$\chi_{\text{in}} = \phi_{\text{in}}\begin{pmatrix} 1 & 0 \\ \varphi & 1 \end{pmatrix}, \qquad \varphi(q) = \int \frac{dk}{\pi} \frac{e^{-it\varepsilon(k)+ikx}s_-(k)s_+(k)}{k-q}, \qquad s_\pm(q) = \frac{1}{\alpha q - \Lambda \pm i}. \tag{S3.17}$$

The corresponding jump $\bar{\mathcal{G}}_{\text{in}}$ in $\phi_{\text{in}}(q)$ across the real line is

$$\bar{\mathcal{G}}_{\text{in}} = \frac{1}{\pi}\begin{pmatrix} \sigma(q)s_- & \sigma(q)e^{-i(t\varepsilon(q)t-qx)} \\ s_-s_+(1-\sigma(q))e^{i(t\varepsilon(q)-qx)} & -\sigma(q)s_+ \end{pmatrix}. \tag{S3.18}$$

To derive this expression we have extensively used that $s_- - s_+ = 2is_-s_+$ and

$$\frac{\varphi_+(q) - \varphi_-(q)}{2\pi i} = \frac{e^{-it\varepsilon(q)+iqx}}{\pi} s_-s_+, \qquad (e_+ - \varphi_\pm e_-)(q) = \sqrt{\sigma(q)}s_\pm(q)\frac{e^{it\varepsilon(q)/2 - iqx/2}}{\pi}. \tag{S3.19}$$

The jump matrix is valid for any distribution $\sigma(q)$. For the thermal distribution it simplifies to

$$\bar{\mathcal{G}}_{\text{in}} = \frac{1}{\pi(e^{\beta\varepsilon(q)}+1)}\begin{pmatrix} s_- & e^{-it\varepsilon(q)+iqx} \\ s_-s_+ e^{i(t-i\beta)\varepsilon(q)-iqx} & -s_+ \end{pmatrix}. \tag{S3.20}$$

For $q \to \infty$, $\phi(q)$ has the same expansion as $\chi(q)$ but with modified potentials $b_{ij}$ and $c_{ij}$. Moreover, from Eq. (S3.17) we conclude the relation between the old and new potentials

$$B_{+-} = b_{+-}, \qquad B_{-+} = b_{-+}, \qquad B_{--} = b_{--}, \qquad B_{++} = b_{++} + h \tag{S3.21}$$

$$C_{+-} = c_{+-} + hb_{--}, \qquad C_{-+} = c_{-+}, \qquad C_{--} = c_{--}, \qquad C_{++} = c_{++} - i\partial_x h + hb_{-+} \tag{S3.22}$$

For the Green's function we then have

$$e^{it\mu}G_{\text{in}} = -b_{++}\tau_{\text{in}}. \tag{S3.23}$$

This expresses the injection Green's function through the potential $b_{++}$ and the $\tau$-function. The potential $b_{++}$ appears from a solution to the RHP with specific jump matrix $\bar{\mathcal{G}}_{\text{in}}$. We will find now a similar representation for the ejection Green's function.

Now let us consider the ejection. We denote the corresponding RHP matrix as $\chi_{\text{ej}}$ and the potentials with the subscripts ej. Following the same procedure as for the injection we obtain

$$e^{-i\mu t} G_{\text{ej}} = B^{\text{ej}}_{--} \tau_{\text{ej}} \tag{S3.24}$$

We perform again the conjugation

$$\chi_{\text{ej}} = \phi_{\text{ej}} \begin{pmatrix} 1 & 0 \\ \varphi_{\text{ej}} & 1 \end{pmatrix}, \qquad \varphi_{\text{ej}}(q) = \int \frac{dk}{\pi} \frac{e^{ikx-it\varepsilon(k)}}{k-q}. \tag{S3.25}$$

The jump matrix $\phi_{\text{ej}}$ for reads

$$\tilde{\mathcal{G}}_{\text{ej}} = \frac{1}{\pi} \begin{pmatrix} -s_+\sigma(q) & s_-s_+\sigma(q)e^{it\varepsilon(q)-iqx} \\ (1-\sigma(q))e^{it\varepsilon(q)-iqx} & \sigma(q)s_- \end{pmatrix}. \tag{S3.26}$$

This jump matrix is structurally similar to $\bar{\mathcal{G}}_{\text{in}}$ from eq. (S3.18) but with $s_+$ and $s_-$ replaced. This can be fixed by performing a further conjugation, $\phi_{\text{ej}} = \sigma_1 \psi_{\text{ej}} \sigma_1$.[1] The resulting jump matrix is

$$\sigma_1 \tilde{\mathcal{G}}_{\text{ej}} \sigma_1 = \frac{1}{\pi} \begin{pmatrix} \sigma(q)s_- & (1-\sigma(q))e^{it\varepsilon(q)-iqx} \\ s_-s_+\sigma(q)e^{it\varepsilon(q)-iqx} & -s_+\sigma(q) \end{pmatrix}. \tag{S3.27}$$

Finally specifying $\sigma(q)$ to be a thermal distribution we conclude the following jump matrix for $\Psi$

$$\sigma_1 \tilde{\mathcal{G}}_{\text{ej}} \sigma_1 = \frac{1}{\pi(e^{\beta\varepsilon(q)}+1)} \begin{pmatrix} s_- & e^{i(t-i\beta)\varepsilon(q)-iqx} \\ s_-s_+ e^{-it\varepsilon(q)+iqx} & -s_+ \end{pmatrix} \tag{S3.28}$$

Comparing this expression with the jump matrix (S3.20) we observe that

$$\bar{\mathcal{G}}_{\text{in}}\Big|_{x,t+i\beta/2} = \sigma_1 \tilde{\mathcal{G}}_{\text{ej}} \sigma_1 \Big|_{-x,-t+i\beta/2}, \tag{S3.29}$$

which leads to the following identity $\psi_{\text{ej}}(-x,-t-i\beta/2) = \phi_{\text{in}}(x,t-i\beta/2)$ and

$$\chi_e(-x,-t-i\beta/2) = \sigma_1 \Phi(x,t-i\beta/2) \sigma_1 \begin{pmatrix} 1 & 0 \\ \tilde{\varphi} & 1 \end{pmatrix}. \tag{S3.30}$$

By comparing the asymptotic expansion of both sides of this equality we conclude that

$$B^{\text{ej}}_{-+}(-x,-t-i\beta/2) = -b_{+-}(x,t-i\beta/2), \quad B^{\text{ej}}_{--}(-x,-t-i\beta/2) = -b_{++}(x,t-i\beta/2), \tag{S3.31}$$

$$C^{\text{ej}}_{-+}(-x,-t-i\beta/2) = -c_{+-}(x,t-i\beta/2). \tag{S3.32}$$

For the ratio of the Green's function we find

$$\frac{G_{\text{ej}}(-x,-t-i\beta/2)}{G_{\text{in}}(x,t-i\beta/2)} = e^{\beta\mu} \frac{\det(1+V_{\text{ej}})}{\det(1+V_{\text{in}})}. \tag{S3.33}$$

We will now prove that the ratio of the determinants is independent of $x$ and $t$. To this end consider derivatives of the kernels $V_{\text{in}}$ and $V_{\text{ej}}$. For the injection, taking into account specific dependence of the kernel, we find

$$\partial_x V(q,q') = \frac{i}{2} \langle E(q)|\sigma_3|E(q)\rangle, \tag{S3.34}$$

$$\partial_t V(q,q') = -\frac{i}{4}(q+q')\langle E(q)|\sigma_3|E(q)\rangle - \frac{ih}{2}\langle E(q)|\sigma_-|E(q)\rangle, \tag{S3.35}$$

with

$$\sigma_- = \begin{pmatrix} 0 & 0 \\ 1 & 0 \end{pmatrix}, \tag{S3.36}$$

---

[1] Matrix $\sigma_1$ appears here on both sides to ensure the same asymptotic behavior of $\phi_{\text{ej}}$ and $\psi_{\text{ej}}$.

and $h$, after the rescaling (S3.5), is given by

$$h = \int_{-\infty}^{\infty} \frac{dq}{\pi} \frac{e^{-it\varepsilon(q)+ixq}}{1+(\alpha q - \Lambda)^2}. \tag{S3.37}$$

The derivatives of the tau-function then read

$$\partial_x \log \tau_{\text{in}} = iB_{+-}, \qquad \partial_t \log \tau_\chi = -i\frac{C_{-+}+C_{+-}}{2} + \frac{ih}{2}B_{--}. \tag{S3.38}$$

and in terms of the potentials of $\phi(q)$ we obtain

$$\partial_x \log \tau_{\text{in}} = ib_{+-}, \qquad \partial_t \log \tau_{\text{in}} = -i(c_{+-}+c_{-+})/2, \tag{S3.39}$$

For the ejection the computations are analogous and we obtain

$$\partial_x \log \tau_{\text{e}} = iB^{\text{ej}}_{+-}, \qquad \partial_t \log \tau_{\text{e}} = -i\frac{C^{\text{ej}}_{-+}+C^{\text{ej}}_{+-}}{2} + \frac{ih_0}{2}B^{\text{ej}}_{--}, \tag{S3.40}$$

with $h_0 = \sqrt{\frac{2}{\pi i t}} e^{ix^2/(2t)}$. If we now specialise the space-time variables in the ejection $\tau$-function to $(-x, -t - i\beta/2)$ and in the injection to $(x, t - i\beta/2)$ we find that both spatial and time derivatives of the two $\tau$-functions are the same. Therefore their ratio is space-time independent. This allows us to conclude that

$$\frac{G_{\text{ej}}(-x, -t - i\beta/2)}{G_{\text{in}}(x, t - i\beta/2)} = e^{\beta\mu} W(\beta, \Lambda). \tag{S3.41}$$

As the last step of deriving the refined detailed balance we compute function $W(\beta, \Lambda)$. Since the ratio of the corresponding determinants does not depend on the coordinates we can evaluate it at $t = 0$ and send $x \to +\infty$. In this limit the simplified kernels can be approximated as

$$V_{i,e}(q,p) \approx \frac{e^{i,e}_+(q)e^{i,e}_-(p) - e^{i,e}_-(q)e^{i,e}_+(q)}{q-p} \tag{S3.42}$$

with

$$e^e_-(q) = \frac{e^{\beta\varepsilon(q)/4-iqx/2}}{\sqrt{\pi}\sqrt{1+e^{\beta\varepsilon(q)}}\sqrt{1+(\alpha q - \Lambda)^2}}, \qquad e^e_+(q) = -\frac{e^{-\beta\varepsilon(q)/4+iqx/2}(\alpha q - \Lambda - i)}{\sqrt{\pi}\sqrt{1+e^{\beta\varepsilon(q)}}\sqrt{1+(\alpha q - \Lambda)^2}} \tag{S3.43}$$

$$e^i_-(q) = \frac{e^{-\beta\varepsilon(q)/4-iqx/2}}{\sqrt{\pi}\sqrt{1+e^{\beta\varepsilon(q)}}(\alpha q - \Lambda + i)}, \qquad e^i_+(q) = -\frac{e^{\beta\varepsilon(q)/4+iqx/2}}{\sqrt{\pi}\sqrt{1+e^{\beta\varepsilon(q)}}}. \tag{S3.44}$$

To study the asymptotic expansion of the determinants we use the method of effective form factors [6]. In this method the exact kernel of the Fredholm determinant is replaced by a simplifed kernel $\bar{V}$ chosen in such a way that both expressions have the same asymptote for large $x$. The simplified kernel is of the Sine-type,

$$\bar{V}(q, q') = \frac{E_+(q)E_-(q') - E_-(q)E_+(q')}{q - q'}, \tag{S3.45}$$

with

$$E_+(q) = \frac{e^{iqx/2+i\pi\nu(q)-g(q)/2}}{\sqrt{\pi}}, \qquad E_-(q) = \frac{e^{-iqx/2+g(q)/2}\sin(\pi\nu(q))}{\sqrt{\pi}}. \tag{S3.46}$$

The asymptotic expansion is then

$$\det(1 + \bar{V}) = \exp\left(\int g(q)\nu'(q)dq\right) \times F[\nu; x]. \tag{S3.47}$$

Functions $g(q)$ and $\nu(q)$ are determined from the original kernels by matching functions $E(q)_\pm(q)$ with $e_\pm(q)$. From this identification we obtain that $\nu(q)$ coincides both for injection and ejection

$$\nu(q) = \nu_e(q) = \nu_i(q) = \frac{1}{2\pi i}\log\left(1 - \frac{2i\sigma(q)}{\alpha q - \Lambda + i}\right), \tag{S3.48}$$

while functions $g$ differ by

$$\delta g(q) = g_e(q) - g_i(q) = \beta\varepsilon(q) + \log\frac{\alpha q - \Lambda + i}{\alpha q - \Lambda - i}. \tag{S3.49}$$

The asymptotic is then given by

$$\lim_{x\to+\infty}\frac{\det(1+V_e)}{\det(1+V_i)} = \exp\left(\int \delta g(q)\nu'(q)dq\right). \tag{S3.50}$$

We can simplify (S3.48), (S3.49) by identically rewriting them as

$$\delta g(q) = \beta\varepsilon(q) - 2i\delta(q), \qquad \nu(q) = \frac{\log(1+e^{-\delta g(q)}) - \log(1+e^{-\beta\varepsilon(q)})}{2\pi i}, \tag{S3.51}$$

where $\delta(k) = \frac{\pi}{2} - \arctan(\Lambda - \alpha k)$ is the Bethe phase shift. Then computation of the integral is straightforward

$$\int \delta g(q)\nu'(q)dq = \beta\int \frac{dq}{\pi}\varepsilon'(q)\sigma(q)\delta(q) = \beta\mathcal{F}_0(\Lambda), \tag{S3.52}$$

with $\mathcal{F}_0$ the intensive contribution to the free energy defined in Eq. (S2.10). This allows us to finally conclude

$$\frac{G_{\text{ej}}(-x,-t-i\beta/2)}{G_{\text{in}}(x,t-i\beta/2)} = e^{\beta\mu+\beta\mathcal{F}_0(\Lambda)}, \tag{S3.53}$$

which completes the derivation of the refined detailed balance.

## S4 Fredholm determinant representation of the Green's functions

In this Section we recall the definition of the Fredholm determinant and provide details on the numerical evaluation of it that we use in our work.

We define the Fredholm determinant of a kernel $\hat{K}$ through an expansion with respect to a parameter $\lambda$

$$\det(1+\lambda\hat{K}) = 1 + \lambda\int dx K(x,x) + \frac{\lambda^2}{2}\int dx_1 dx_2 \det\begin{pmatrix} K(x_1,x_1) & K(x_1,x_2) \\ K(x_2,x_1) & K(x_2,x_2) \end{pmatrix}$$
$$+ \frac{\lambda^3}{3!}\int\prod_{j=1}^{3}dx_j \det\begin{pmatrix} K(x_1,x_1) & K(x_1,x_2) & K(x_1,x_3) \\ K(x_2,x_1) & K(x_2,x_2) & K(x_2,x_3) \\ K(x_3,x_1) & K(x_3,x_2) & K(x_3,x_3) \end{pmatrix} + \ldots, \tag{S4.1}$$

with the integration range extending from $-\infty$ to $+\infty$. This definition can be easily generalized to include an additional measure $\sigma(x)$. We assume that $0 < \sigma(x) < 1$ and vanishing quickly enough to guarantee the convergence of the integrals. It suffices for the kernel to be of trace-class condition

$$\left|\int dx\,\sigma(x)K(x,x)\right| < \infty. \tag{S4.2}$$

Then

$$\det_\sigma(1+\lambda\hat{K}) = \det(1+\lambda(\sqrt{\sigma}\hat{K}\sqrt{\sigma})), \tag{S4.3}$$

where $(\sqrt{\sigma}\hat{K}\sqrt{\sigma})(x,x') = \sqrt{\sigma(x)}K(x,x')\sqrt{\sigma(x')}$. For a numerical evaluation of the Fredholm determinant it is practical to use the quadrature at the level of the determinant rather then its expansion. This approximates the Fredholm determinant by a determinant of a finite dimensional matrix,

$$\det(1+\hat{K}) \approx \det_N\left(\delta_{jk} + \sqrt{\Delta x_j}K(x_j,x_k)\sqrt{\Delta x_k}\right), \tag{S4.4}$$

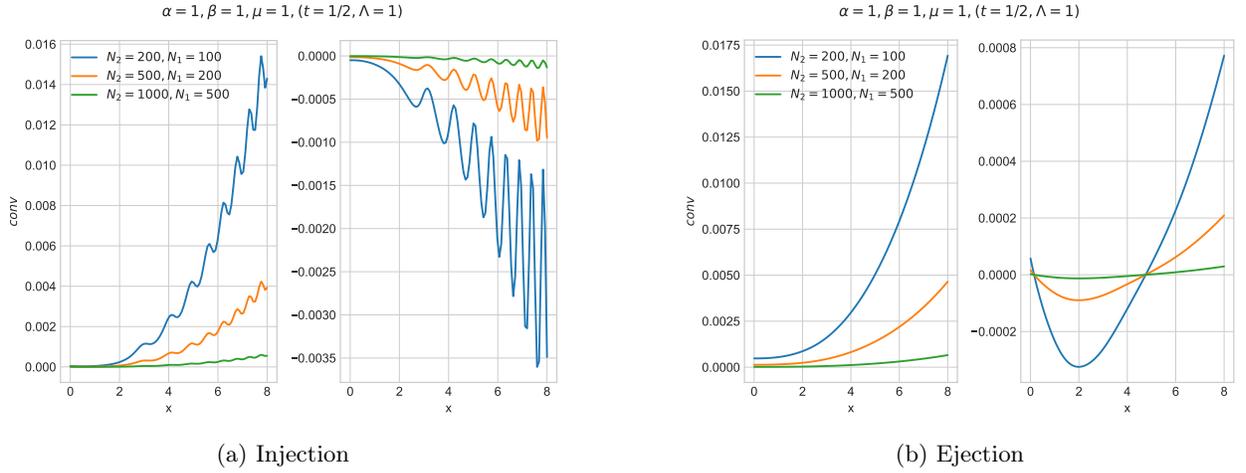

(a) Injection

(b) Ejection

FIG. S1: Plots of the convergence measure (S4.5). We plot real and imaginary parts of the two Green's function for the parameters of the system shown above the plots. The results show that increasing the size of the grid from $N = 500$ to $N = 1000$ points changes the values of the two functions by at most 0.1%.

with $\{x_j\}_{j=1}^N$ the centers of a discrete grid consisting of $N$ points with spacing $\{\Delta x_j\}_{j=1}^N$. This formula is easily adaptable to include the meassure $\sigma(x)$.

The kernels are of a Sine-type and are of the form $(f(x) - f(x'))/(x - x')$. To avoid the problematic $x = x'$ point we discretize $x$ and $x'$ on grids shifted with respect to each other such that $x_j \neq x'_k$ for any $j$ and $k$.

To establish the convergence of the results we evaluate the determinants on grids with different number of points. To quantify the convergence we define

$$\text{conv} = \frac{G^{(N_2)}(x, t, \Lambda) - G^{(N_1)}(x, t, \Lambda)}{G^{(N_1)}(x, t, \Lambda)}, \tag{S4.5}$$

where $G^{(N)}(x, t, \Lambda)$ is the Green's function (either injection or ejection) computed on a grid consisting of $N$ points. The results for the grids of $N = 100, 200, 500, 1000$ points and for both correlators are shown in fig. S1. We have verified that the fast convergence rate holds also for other values of the parameters of the system.

[1] O. Gamayun, A. G. Pronko, and M. B. Zvonarev, Nuclear Physics B **892**, 83 (2015).
[2] O. Gamayun, A. G. Pronko, and M. B. Zvonarev, New Journal of Physics **18**, 045005 (2016).
[3] T. D. Lee, F. E. Low, and D. Pines, Phys. Rev. **90**, 297 (1953).
[4] D. M. Edwards, Progress of Theoretical Physics Supplement **101**, 453 (1990), https://academic.oup.com/ptps/article-pdf/doi/10.1143/PTP.101.453/5454680/101-453.pdf.
[5] H. Castella and X. Zotos, Phys. Rev. B **47**, 16186 (1993).
[6] O. Gamayun, M. Panfil, and F. T. Sant'Ana, Phys. Rev. A **106**, 023305 (2022).
[7] V. E. Korepin, N. M. Bogoliubov, and A. G. Izergin, *Quantum Inverse Scattering Method and Correlation Functions* (Cambridge University Press, 1993).
[8] K. K. Kozlowski, Advances in Theoretical and Mathematical Physics **15**, 1655 (2011).